\documentclass[preprint2]{aastex}

\title{Discovery of twin kHz QPOs in the peculiar X-ray binary Circinus X-1}

\author{S. Boutloukos\altaffilmark{1,2},
M. van der Klis\altaffilmark{1},\\
D. Altamirano\altaffilmark{1},
M. Klein-Wolt\altaffilmark{1},
R. Wijnands\altaffilmark{1},
P.G. Jonker\altaffilmark{3,4,5},
R.P. Fender\altaffilmark{1,6}}

\email{stratos@science.uva.nl}

\altaffiltext{1}{Astronomical Institute Anton Pannekoek,
University of Amsterdam, and Center for High Energy Astrophysics,
Kruislaan 403, 1098 SJ Amsterdam, The Netherlands.}
\altaffiltext{2}{Theoretical Astrophysics, University of T\"ubingen, Auf der
Morgenstelle 10, 72076 T\"ubingen, Germany}
\altaffiltext{3}{SRON, Netherlands Institute for Space Research,
Sorbonnelaan 2, 3584 CA, Utrecht, The Netherlands.}
\altaffiltext{4}{Harvard--Smithsonian  Center for Astrophysics, 60 Garden Street, 
Cambridge, MA~02138, Massachusetts, U.S.A.}
\altaffiltext{5}{Astronomical Institute, Utrecht University, P.O.Box 80000, 3508 TA, 
Utrecht, The Netherlands}
\altaffiltext{6}{School of Physics and Astronomy, University of
Southampton, Southampton SO17 1BJ, United Kingdom}

\begin{document}

\begin{abstract}

We report the discovery with the RXTE/PCA of twin kHz QPOs in the
peculiar X-ray binary Circinus X-1. At eleven different epochs we
observed two simultaneous kHz QPOs with centroid frequencies of up to
$\sim$225 and $\sim$500~Hz and significances of up to 6.3 and
5.5$\sigma$ for the lower and upper kHz QPO, respectively.  The
frequency range over which the twin kHz QPOs are seen is for the most
part well below that of other sources, and extends down to centroid
frequencies of $\sim$56~Hz and $\sim$230~Hz respectively for the lower
and simultaneously observed upper kHz QPO.  The strongly variable QPO
frequencies and their observed correlations clearly indicate that the
twin peaks are the kHz QPOs such as typically seen from low-magnetic
field neutron stars, and not black-hole high-frequency QPOs.  This
confirms that Cir~X-1 is a neutron star, as suspected since the
detection of Type-I X-ray bursts from the field of the source 20 years
ago.  The kHz QPO peak separation varies over the largest range yet
seen, $\sim$175--340~Hz, and {\it increases} as a function of kHz QPO
frequency.  This is contrary to what has been observed in other
sources but in good agreement with predictions of the relativistic
precession model and Alfv\'en wave models at these low QPO
frequencies.  Current beat-frequency models would require further
modification in order to accommodate this.  
 A total of 67 observations showed only a lower kHz QPO; this 
QPO can be followed down all the way to a centroid frequency of
$\sim$12~Hz ($Q\sim$0.5). Two observations showed only an upper kHz
QPO near 450~Hz. In addition, a strong low-frequency quasi-periodic
oscillation (LF QPO) is seen in the 1--30~Hz range, as well as two
components above and below this QPO.  The frequency-frequency
correlations between the kHz QPOs and the LF QPO are in good agreement
with those found previously in Z sources, confirming that Cir~X-1 may
be a peculiar Z source, and extend these correlations to frequencies a
factor 1.5--2.3 lower.  We suggest that the low frequency range over
which the kHz QPOs occur in Cir~X-1 and to a lesser extent in (other)
Z sources, might be due to a stronger radial accretion flow relative
to the disk flow than in other kHz QPO sources, possibly related to
the nature of the companion star.
\end{abstract}

\section{Introduction}\label{intro} 
Low-mass X-ray binaries (LMXBs) often exhibit quasi-periodic
oscillations (QPOs) in their X-ray flux (\citealt{review} for a
review).  The properties of these QPOs vary as a function of the
location of the source in the track it traces out in X-ray color-color
and hardness-intensity diagrams (CD and HIDs).  Among the low
magnetic-field neutron-star systems, systematic differences in the
CD/HID tracks and in the associated rapid X-ray variability allow to
distinguish two subclasses: Z and atoll sources, named after the
shapes of their tracks \citep{hk89}. Both in Z sources and in atoll
sources, the QPO frequency tends to increase in the same direction as the mass
accretion rate is inferred to be increasing, at least on short time
scales (see \citealt{parlines}).

In about 20 LMXBs containing a low-magnetic field neutron star, two
kHz QPOs have been seen simultaneously (twin kHz QPOs) with a centroid
frequency separation $\Delta\nu$ of a few hundred~Hz that decreases
typically by a few tens of~Hz when the QPO frequencies themselves
increase by hundreds of~Hz (roughly, from 200 to 900~Hz and from 500
to 1200~Hz, respectively).  If the spin frequency of the neutron star,
$\nu_s$, is known, then $\Delta\nu$ has always been found to be either
roughly equal to $\nu_s$, or to $\nu_s/2$, depending on the
source. The frequencies of the twin kHz QPOs (called the lower and the
upper kHz QPO, or $L_\ell$ and $L_u$, in order of increasing
frequency) are tightly correlated to each other, following a relation
that is the same across all sources \citep{psalt98,pbvdk,steve}.  Both
kHz QPO frequencies also (e.g., \citealt{pbvdk}) correlate with the frequencies of other
variability components such as, in atoll sources, the low-frequency
quasi-periodic oscillation (LF QPO; a sharp QPO peak in the 1--60~Hz
range designated $L_{LF}$ -- cf. \citealt{steve} for some of the
terminology) and the break component (a broad noise component with a
frequency in the 0.01--20~Hz range designated $L_b$).  The components
most similar to these in the Z sources (and suspected to be physically
the same) are the horizontal branch oscillation (HBO) and the low
frequency noise (LFN), but the correlations of the frequencies of
these components with those of the kHz QPOs differ somewhat
\citep{straaten}.

Twin kHz QPOs are a neutron-star signature.  In some LMXBs containing
a black hole, high-frequency quasi-periodic oscillations (HF QPOs)
have been observed. These are different from kHz QPOs: they tend to
recur near fixed frequencies that are different in each source, and
when more than one HF QPO occurs in the same source the QPOs usually
appear at different times and at frequencies often roughly consistent with a
fixed 2:3 ratio \citep{stroh,mccrem}.

Numerous models have been proposed to explain the nature of the kHz
QPOs and describe their correlations.  Some refer to a blob orbiting
inside the accretion disk, either self-luminous or scattering the
X-rays, or producing a hot spot on the neutron-star surface by the
infall of matter from the blob while it is orbiting at the inner edge
of the disk.  These models produce a rotating pattern of radiation
causing a so-called 'beaming' modulation of the observed flux.  The
orbital frequency of a particle in an equatorial circular orbit around
a point mass with mass $M$ and angular momentum $J$ (approximating the
spacetime metric around the neutron star by Kerr spacetime) is
$\nu_{\phi}=\sqrt{GM/(4\pi^2r^3)} / (1+j(r_g/r)^{3/2})$, where
$r_g\equiv GM/c^2$ and $j\equiv Jc/GM^2$. The corresponding radial
epicyclic frequency in a slightly eccentric orbit and for negligible
frame dragging ($j\approx0$) is
$\nu_r\approx\nu_{\phi}\sqrt{1-6r_g/r}$.

In the relativistic precession model \citep{stella} as well as in the
sonic-point spin-resonance model \citep{cole} the upper kHz QPO $L_u$
is identified with $\nu_{\phi}$. The lower kHz QPO $L_\ell$ is
interpreted in the first case as due to the periastron precession
frequency $\nu_{\phi}-\nu_r$, and in the second case as a beat between
$\nu_{\phi}$ at the sonic radius and the spin frequency (at
approximately $\nu_{\phi}-\nu_{spin}$ or $\nu_{\phi}-\nu_{spin}/2$
depending on conditions at the spin-resonance radius where the beat
emission is generated). \citet{23,1235} suggest that QPOs occurring in
neutron stars are similar to those seen in black holes and result from
resonance frequencies in the disk, but in neutron stars are tuned away
from the fixed frequencies seen in black holes by additional physical
effects (but see also \citealt{distribution}).  Models related to
orbital motion are in general more predictive of the frequencies of
the QPOs than of their amplitudes and coherences, as constructing a
model for the modulation mechanism or describing the damping of the
oscillations is more difficult than explaining these frequencies as a
(combination of) general-relativistic orbital and epicyclic
frequencies.  A maximum frequency according to such models would be
the orbital frequency at the innermost (marginally) stable circular
orbit (ISCO) of the compact object $\nu_{ISCO}\approx
(c^3/2\pi6^{3/2}GM)(1+0.75j)$ for Kerr spacetime or, under certain
conditions, the orbital frequency at the marginally bound orbital
radius.  Both these frequencies scale inversely with mass.  For an
extensive review on timing features of LMXBs and QPO models see
\citet{review}.

Circinus X-1 is a galactic X-ray source located 
in the galactic plane at a distance that has
variously been reported to lie in the range 4--12 kpc
\citep{austr,glass,jonknele,iaria}.  The system is thought to have a
17-d highly eccentric ($e\sim0.8$) orbit leading to periodically
enhanced mass transfer at periastron \citep{murdin}.  From optical and
infrared observations a sub-giant companion star of 3--5~$M_{\sun}$
was suggested \citep{mass}. More recently, \citet{jonker06} found
evidence for supergiant characteristics in the optical spectrum of the
system which might indicate an even higher companion mass. So, Cir~X-1
is formally a high-mass X-ray binary.  However, its X-ray
phenomenology corresponds to that of a low magnetic-field compact
object such as more commonly found in LMXBs.  The strong radio
emission \citep[see, e.g.,][]{radio}, highly relativistic jets
\citep{jets}, hard X-ray emission \citep{iaria01,ding} and very strong
X-ray variability \citep{jones, samimi, atoll}, resemble properties
generally more typical of black holes.  However, type-I X-ray bursts
seen in 1986 from the field of the source and most likely originating
in Cir~X-1 itself \citep{flashes} classified it as a likely neutron
star, with later spectroscopic analysis \citep{tueb,gierl} supporting
this at least at a phenomenological level.  No bursts were reported
from the source since 1986.  Low-frequency noise, a clear
low-frequency QPO $L_{LF}$ and a broad feature at 100--200~Hz were
seen with EXOSAT \citep{tennant87}. At low flux levels Cir~X-1
exhibited atoll-source behavior \citep{atoll} including strong
band-limited noise, while subsequent observations at higher flux on
occasion, and in addition to a wide variety of non-Z-like patterns,
showed a Z-source-like track in the HID \citep{shirey}.  On the
horizontal branch of this track, \citet{shireyI,shireyII} found the LF
QPO to move between 1.3~Hz and 32 Hz, in strong correlation with the
break frequency at lower frequencies and the frequency of the broad
high-frequency feature, now seen to range from 20--100
Hz. \citet{pbvdk} interpreted this broad peak as a low-frequency,
low-coherence version of the lower kHz QPO $L_\ell$ as seen in Z and
atoll sources.  They showed that its frequency fits on the relation
seen in those sources between the frequencies of $L_\ell$ and
$L_{LF}$, extending it down to an $L_\ell$ centroid frequency of
$\sim$20~Hz and, interestingly, providing a link with black-hole
phenomenology.  No twin kHz QPOs were reported from Cir~X-1 up to now.

In this paper, we concentrate on the high-frequency part of the power
spectra of Cir~X-1 and perform an extensive survey for kHz QPOs. We
find that twin kHz QPOs do in fact occur in Cir~X-1.  We describe our
methodology in Section~\ref{method}, present our results in
Section~\ref{results} and comment on consequences for phenomenological
systematics and theoretical models in Section~\ref{dis}.

\section{Observations and Analysis}\label{method} 
We used the data on Cir~X-1 obtained with the proportional counter
array \citep[PCA;][]{jahoda} on board the Rossi X-ray Timing Explorer
(RXTE) from March 1996 till January 2005 listed in Table~\ref{obsids}.
This includes all data publicly available in February 2005 as well as
the then proprietary data set 90025, and comprises about 2 Ms of data
distributed over 21 RXTE programs in 9 RXTE Observing Cycles: a
total of 497 RXTE observations each with typically 3--6 ks of useful
data and identified by a unique RXTE Observation ID.

The source was observed with 13 different combinations of instrumental
modes covering the full energy range (channels 0--249, effective
energy range 3--60~keV), and one more starting from channel 8
(3.7~keV) with which 21 observations were taken in 2003 (Cycle~8),
when the intensity was low (below 0.4~Crab) and from which we
eventually did not report any power spectra.  We divided each
observation into continuous time segments of 16, 64, 128 or 256 s,
depending on the time resolution such that they contained 2$^{20}$
points. The count rates from all of the PCA's available Proportional
Counter Units (PCUs) were summed and Fourier-transformed. The
resulting power spectra were averaged to give one average power
spectrum per observation.

Most of our power spectra (378 of the 497) either showed no
appreciable broad-band power at all or only a broad plateau-like
feature at low frequency ($\la10$~Hz).  These power spectra were
inspected visually for high-frequency features, if necessary followed
by fitting using the procedures described below in order to determine
the significance of any features.  Taking into account the number of
trials inherent in this search, no significant high-frequency features
were found.  The most significant feature (at $\sim$721~Hz and with
$Q$$\sim$8.5) had a single-trial significance of 3.8$\sigma$.
More complex power spectra, containing a significant and narrow
($Q\ga$1) LF QPO in the 1--50~Hz range and at least one more feature
described well by a Lorentzian, were seen in the remaining
observations.  We performed a full analysis of these 119 power spectra
using the multi-Lorentzian fit method described below.

Since this work concentrates on kHz QPOs, careful subtraction of the
deadtime-modified Poisson-noise power is required. A full description
of our approach is provided in Appendix~\ref{appendix}.  We used the
expression proposed by \citet{zhang,zhang2} to describe the Poisson
spectrum, but with values of the event deadtime and VLE window of
$t_d=8.87\mu s$ and $t_{VLE}=162 \mu s$ based on direct fits of that
model to our data.  As explained in the Appendix, this method is
slightly more conservative with respect to the
significances of the twin kHz QPOs than the other methods used in
previous work, and provides a consistent estimate of the
deadtime-modified Poisson-noise power spectrum throughout our data.

After subtracting the noise, we renormalized the power spectra to
source squared fractional rms amplitude and fitted them using standard
chi-squared methods.  We used a multi-Lorentzian fit function in the
($\nu_{max}$, $Q$) representation \citep{vmax}, where the so-called
characteristic frequency $\nu_{max}$ of the Lorentzian is given by
$\nu_{max} \equiv \sqrt{\nu_0^2+\Delta^2}$ with $\nu_0$ the
Lorentzian's centroid frequency and $\Delta$ its HWHM, and where its
quality factor is $Q\equiv\nu_0/2\Delta$.  A power law was added at
low frequencies when needed. The fits were made without any reference
to expected features, and include all significant ($>3\sigma$; all
significances quoted herein are single-trial) Lorentzians that could
be fitted.  All errors quoted are 1-$\sigma$ single-parameter
($\Delta\chi^2=1$).

To recover the Lorentzian centroid frequency $\nu_0$, we used the
expression $\nu_0=\nu_{max}/\sqrt{1+{1/{4Q^2}}}$.  Since all
Lorentzians for which we applied this conversion are rather
sharp-peaked ($Q\ga 1$), the errors computed for $\nu_{max}$ are
approximately the same as those for $\nu_0$.  For the twin kHz QPOs,
we explicitly re-estimated the errors by refitting in
centroid-frequency representation, and typically found the errors to
differ by $<$5\% (usually $<$1~Hz), and negligible differences in the
parameters themselves.  For the other fits we adopted the $\nu_{max}$
errors also for the centroid frequencies.  In the following, all
frequencies quoted are in terms of the characteristic frequency
$\nu_{max}$ unless otherwise noted.  Subscripts denote the variability
component ($\nu_u$ for upper kHz QPO, etc.); centroid frequencies are
denoted with superscript $0$ as follows: $\nu_u^0$.

We used the 16-s time-resolution Standard-2 mode to calculate X-ray
colors as in \citet{diego} but with bands as defined by
\citet{shirey}.  Dropouts were removed, background was subtracted and
deadtime corrections were made.  All active PCUs were used for this
analysis.  The energy bands chosen for colors and intensity are given
in the captions to Figs.~\ref{ccd} and \ref{hid} -- energy channels
were interpolated to approximate the exact same bands in all data
sets.  In order to correct for the gain changes as well as the
differences in effective area between the PCUs, we normalized all
color and intensity estimates to the corresponding Crab values
\citep{kuulkers,straaten} that are close in time but in the same RXTE
gain epoch.

%soft = 4.8-3.4 / 3.4-2 keV
%hard = 18.-8.5 / 8.5-4.8 keV
%broad = 13-6.3 / 6.3 - 2 keV
%Intensity = 2-18 keV

\section{Results}\label{results}
In 80 of the 119 fitted power spectra we found a QPO which we interpret
as a kHz QPO (see below). In 7 of these we found a significant
($>3\sigma$) second kHz QPO. A second kHz QPO with a significance
between 2.5 and 3$\sigma$ was found in 4 more cases.  We discuss
whether these are real or not in Section \ref{twin}.

The best power spectrum exhibiting twin kHz QPOs (data set H;
observation 20094-01-01-01) is shown in Fig.~\ref{ps}.  Twin kHz
QPOs, 6.3 and 5.5$\sigma$ significant, are clearly seen near 140 and
460~Hz, respectively, as well as a strong LF QPO near 10~Hz, broad
noise, and a power law at low frequency.  The power spectra of all 11
observations with twin kHz QPOs are shown in Fig.~\ref{fits} together
with all fitted components.  The prominent narrow feature peaking
somewhere between 1 and 30~Hz is the LF QPO $L_{LF}$.  It is seen in
all 119 observations, sometimes as a double peak, which may be due to
averaging over a moving feature.  When due to this it was necessary to
fit the LF QPO with two Lorentzians, we adopted the frequency of the
most significant (usually, the lower-frequency) of the two as that of
the LF QPO in further analysis. A low-frequency noise or break
component was seen below $\nu_{LF}$ in all but three cases while an
additional component was detected at or above $\nu_{LF}$ in more than
half (78) of the power spectra. The nature of this component is
further discussed in Section~\ref{other}. Power laws were needed in
only nine cases, and always only produced appreciable power below
$\sim$1~Hz.

The average X-ray colors of each observation with kHz QPOs are plotted
on top of a 16-s time-resolution color-color diagram (CD) of all
Cir~X-1 data in Fig.~\ref{ccd}.  The observations with twin kHz QPOs
and single kHz QPOs are set out by different symbols, as are the 16-s
data already reported by \citet{shirey}.  Twin and single kHz QPOs
occur over a range of colors; the frequency of the lower kHz QPO
tends to increase towards both lower and higher colors, but the
correlations are weak.  A hardness-intensity diagram following
the same conventions is provided in Fig.~\ref{hid}.  Most of the
observations showing twin kHz QPOs were within a quite narrow band of
intensities between 2545 and 2770 c/s/PCU, or
about 1.0--1.1~Crab (except data set H at 1991 c/s/PCU or
$\sim$0.8 Crab, which is also the overall hardest kHz QPO
observation; the uppermost black triangle in Figs.~\ref{ccd} and
\ref{hid}).  Presumably related to this, we find no clear correlation
between kHz QPO frequency and intensity, although there is some weak
evidence for parallel tracks in frequency vs. intensity, as has been
seen in other sources.  Most kHz QPOs, and all twins, are found well
above the Z tracks identified by \citet{shirey} which are confined to
Crab-normalized broad color $\la0.6$, i.e., most QPOs are associated
with a considerably harder overall spectrum than that in the Z tracks.
Fig.~\ref{lc} shows a long term light curve of the source where each
data point is the average of one observation.  It can be clearly seen
that most of the observations with kHz QPOs, and all with twin kHz
QPOs, occurred at relatively early epochs, when the average source
flux was higher than later in time, and that most cluster near
intensities of 1~Crab.

To investigate the relations between the frequencies of the
power-spectral components we plot the characteristic frequency of all
features vs. $\nu_{LF}$ in Fig.~\ref{fr-fr}. The points can be easily
separated into five groups of correlated frequencies. The LF QPO ({\it
small filled circles}) is the most significant feature in our power
spectra.  A dispersed group of points below the LF QPO ({\it
triangles}) corresponds to the low-frequency noise or break component
(see also Section~3.3). At a frequency between 1 and 3 times that of
the LF QPO we see a group of points associated with a component ({\it
open circles}) that is usually dominated by a harmonic to the LF QPO
(see Section~\ref{other}).  Parallel to that and at somewhat higher
frequencies there is an almost continuous series of points extending
from $\sim$18 up to $\sim$380~Hz ({\it large filed circles}).  Yet
another group of points above these has similar characteristics,
although less dense, and extends from $\sim$235 up to $\sim$580~Hz
({\it squares}). From the high frequencies to which they extend, and
their frequency variability over more than an order of magnitude, we
identify these latter two components as kHz QPOs.  We conclude that
the highest-frequency group ({\it squares}) corresponds to the upper
kHz QPO ($L_u$) and the group below that ({\it large filled circles})
to the lower kHz QPO ($L_\ell$).  These $L_\ell$ components coincide
with the lower kHz QPOs as tentatively identified by \citet[][see
Section~1]{pbvdk}, and are found here over a wider frequency range.
Note that in
accordance with previous work we use the term kHz QPOs for components
that can have high frequencies, but are also seen down to a
characteristic frequency as low as (in our case) 18~Hz.

\subsection{Twin kHz QPOs}\label{twin}
As noted in Section~\ref{results}, Fig.~\ref{fits} shows the power
spectra of the observations exhibiting twin kHz QPOs that were both
significant at $>$3$\sigma$ (up to 17$\sigma$ for $L_\ell$ and up to
5.5$\sigma$ for $L_u$), as well as those with a significance
$>$3$\sigma$ for one and $>$2.5$\sigma$ for the other.  From
Figs.~\ref{ccd} and \ref{hid} we see that the points of the
2.5--3$\sigma$ group fall right among the $>$3$\sigma$ group.  The QPO
frequencies of the two groups are also in the same range, and the
frequency correlations seem to hold for both groups of points
(Fig.~\ref{fr-fr}).  We conclude that it is safe to include the four
observations with a kHz QPO between 2.5 and 3$\sigma$ in our sample of
twin kHz QPOs; of course they have somewhat larger statistical error
bars. Weak features above $\nu_u$ visible in Fig.~\ref{fits} (e.g. in
power spectrum E), and in between $\nu_u$ and $\nu_\ell$ (e.g. in
power spectrum G) are not significant.
We remain thus with 11 twin kHz QPOs. Table \ref{khz-table} lists
their best-fit centroid frequencies, coherences and fractional rms
amplitudes, as well as the peak centroid separation $\Delta\nu$ and
the centroid frequency ratio.  Note that $\Delta\nu$ tends to {\it
increase} with QPO frequency and that the frequency ratio is
inconsistent with 2:3.  This is further discussed in Section~4.

\subsection{Single kHz QPOs}{\label{single} 

Apart from the 11 twin kHz QPOs we measured an additional 69 single
ones.  Their frequency shows a strong correlation with that of
$L_{LF}$.  As can be seen in Fig.~\ref{fr-fr}, these single kHz QPOs
are nearly all consistent with being lower kHz QPOs: the 11 lower kHz
QPOs among the twin peaks are clearly part of the same group of
points.  Only the two highest-frequency single kHz QPO points
(20094-01-01-00 and 20094-01-01-000) seem to diverge from this
correlation (particularly in terms of centroid frequencies, not
shown).  Since these points seem instead to extend the relation for
the upper kHz QPOs, and because a less significant feature is present
in those power spectra at frequencies consistent with a lower kHz QPO,
we classify these two QPOs as $L_u$, with $L_\ell$ 
remaining formally undetected due to statistics.  This is consistent
with the decrease in lower kHz QPO amplitude with increasing frequency
(Fig.~\ref{rms}).  The best-fit power law for the lower kHz QPO points
is $\nu_\ell=(19.3\pm0.6)\times\nu_{LF}^{0.77\pm0.02}$~Hz, with a
$\chi^2/dof$ of 17.8 for 76 degrees of freedom, clearly indicating the
existence of non-statistical scatter in these frequencies; here and
below the error bars quoted for power-law fit parameters were
multiplied by $\sqrt{\chi^2/dof}$ to take account of this scatter. For
$L_u$, we similarly find $\nu_u=(130\pm30) \times
\nu_{LF}^{0.50\pm0.09}$~Hz; $\chi^2/dof=3.4$ for 11 degrees of
freedom.

The quality factors of the kHz QPOs ranged up to $Q$$\sim$6 for
$L_\ell$ and up to $Q$$\sim$2.5 for $L_u$ but were more often around
1.  No clear trend was detected in the $Q$ values as a function of QPO
frequency.  Averaging over a moving feature may in some cases have
broadened the peak somewhat, but experiments with data subsets show
this effect to be minor down to accessible time scales
($\sim$5000~s). Only in two cases evidence was detected of a
deviation from the standard Lorentzian profile strong enough to allow
a significant additional component to be fitted.  We used the results
of a single Lorentzian fit also in these two cases.  
 
Fig.~\ref{rms} shows the fractional rms amplitude of each kHz QPO as a
function of its own characteristic frequency.  Some general trends
are seen, where the rms of $L_\ell$ monotonically decreases (but more
rapidly below a $\nu_\ell$ of 200~Hz than above that frequency) and
the rms of $L_u$ varies much less, but may first rise and then fall
again. This is somewhat reminiscent of the amplitude variation of the
kHz QPOs in the Z source GX~5$-$1 \citep{jonker} and different from
what has been observed in a number of atoll sources \citep{mendezrms}.
Both QPOs appear to vary in amplitude gradually and tend to become
undetectable when they fall below an rms amplitude of $\sim$1.5\%.

\subsection{Other power spectral components}\label{other} 

The lower frequency power spectral components are not the prime
objective of this work, but we briefly report here on a number of
immediate findings concerning these components that emerge from our
analysis.  

The component above $L_{LF}$ often is rather coherent (with
Q up to $\sim$8), and when it is (for $Q\ga2$), it has a centroid
frequency twice that of the LF QPO (Fig.~\ref{twoLF}) to good
precision, indicating that mostly this component is the harmonic
$L_{2LF}$ of the LF QPO.  However, when it is less coherent the
component often appears to be dominated by broad structure around or
just above the LF QPO.  It can then have frequencies down to a
characteristic frequency as low as that of $L_{LF}$ (see
Fig.~\ref{fr-fr}), and to well below that component in terms of
centroid frequency (see Fig.~\ref{twoLF}).

Although an increase in the break frequency of the low-frequency noise
with $\nu_{LF}$ is evident in Fig.~\ref{fr-fr}, the scatter is
large. The power spectra at low frequencies may be more complex than
allowed in our fits, which mainly concentrate on kHz QPOs. Possibly,
two parallel groups of points are present, indicating two break
frequency components as observed in other sources \citep[$L_b$ and
$L_{b2}$; cf.][]{straaten}.  This seems for example to be the case in
power spectrum F. More work is needed to resolve this issue.

\section{Discussion}\label{dis}  
We have discovered twin kHz QPOs in Cir~X-1 in 11 separate RXTE/PCA
observations.  Among these 11 observations the centroid frequency of
the upper kHz QPO ranged from 229$\pm$18 to 505$\pm$51~Hz and that of
the lower kHz QPO from 56.1$\pm$1.3 to 226$\pm$18~Hz.  The
lowest-frequency twin kHz QPOs observed so far, in the accreting
millisecond pulsar XTE~J1807$-$294, had $\sim$127 and $\sim$353~Hz centroid
frequencies \citep{linares}; our Cir~X-1 data extend to well below this.
A total of 80 observations showed at least one kHz QPO, with centroid
frequencies extending down to as low as 12.5~Hz.  The centroid
frequency separation of the twin kHz QPOs varies by a factor of two,
spanning a range, 173$\pm$18 to 340$\pm$64~Hz, similar to that covered
by all previous twin kHz QPO detections in other sources taken
together.

In order to further compare the kHz QPOs in Cir~X-1 with those in
other sources we examine the locus of these QPOs in two
frequency-frequency correlation diagrams that have been considered in
previous works.  In Fig.~\ref{pbk} we plot $\nu_u^0$ and $\nu_{LF}^0$
vs. $\nu_\ell^0$ in the \citet{pbvdk} diagram containing data for both
neutron stars and black holes.  The lower kHz QPO points of Cir~X-1
had previously been used by these authors to suggest a link between
the lower kHz QPO in Z and atoll sources above a $\nu_\ell^0$ of
200~Hz, and a broad feature in black holes and weak X-ray burst
sources below a $\nu_\ell^0$ of 10~Hz (both plotted in Fig.~\ref{pbk}
using {\it open circles}).  This latter broad feature was dubbed
$L_{\ell ow}$ by \citet{steve}, who based on the relations of
amplitude and coherence with frequency, and on systematic shifts in
the frequency-frequency correlations observed in some millisecond
pulsars questioned the identification with $L_\ell$
\citep{straaten, steve}.  Our identification of the twin kHz QPOs in
Cir~X-1 confirms that at least in Cir~X-1 this feature is indeed the
lower kHz QPO as proposed by \citet{pbvdk}.  Our newly discovered
upper kHz QPO points smoothly extend the $\nu_u^0$ vs. $\nu_\ell^0$
relation to frequencies a factor 4 lower than originally reported
(\citealt{pbvdk}; upper track in Fig.~\ref{pbk}).  The best-fit power
law to all points including Cir~X-1 is $\nu_u^0=(23\pm3) \times
{\nu_\ell^0}^{(0.57\pm0.02)}$~Hz ($\chi^2/dof=20$ for 82 degrees of
freedom).  As can be seen in Fig.~\ref{pbk}, this relation provides an
excellent match to Cir~X-1 as well as to the data on the other
sources.

In Fig.~\ref{vstrfig} we plot the characteristic frequencies of all
other power spectral components in the 11 twin kHz QPO
power spectra vs. $\nu_u$ in a diagram originating with \citet{multilor}
containing data for atoll sources, weak burst sources and Z sources.
In this diagram it can be seen that the $\nu_\ell$ vs. $\nu_u$
relation, as smoothly extended with our Cir~X-1 points, does {\it not}
connect to the $\nu_{\ell ow}$ vs. $\nu_u$ relation, supporting van
Straaten et al.'s suspicion that $L_\ell$ and $L_{\ell ow}$ may be
different components.  Further work on the other sources contributing
to these correlations is required to definitively resolve this issue;
this is beyond the scope of this paper. 

The large range over which the twin kHz QPO frequencies vary, the
strong deviation from a 2:3 ratio and the frequent simultaneous
appearance of the two peaks are very different from what is seen in
black hole high-frequency QPOs.  Instead, the way in which the kHz QPO
frequencies correlate to each other and to the frequencies of other
variability components is consistent with what is seen in neutron
stars.  These observations therefore support the identification of
Cir~X-1 as a neutron star (cf. Section~1).  In Fig.~\ref{vstrfig} it
can be seen that the LF QPO in Cir~X-1 fits well on the $L_h$ track in
this diagram, but the low-frequency noise or break component does not
match well to the $L_b$ track.  Cir~X-1 shares both these
characteristics with the Z sources \citep{straaten}, giving further
support to the proposal of \citet{shirey} that Cir~X-1 is a (peculiar)
Z source.  The low coherence, relative weakness, and low frequency
range of our twin kHz QPOs as well as the detection of a LF QPO
harmonic (such as also commonly seen in Z source HBOs) are also in
accordance with a Z source classification.  However, the band-limited
noise can be much stronger \citep{atoll}, and the kHz QPOs can reach
much lower frequencies in Cir~X-1 than has been observed in any
(other) Z source.

The extended range of kHz QPO frequencies in Cir~X-1 allows unique
tests of theoretical models.  Perhaps most strikingly, the increase in
the frequency separation $\Delta\nu$ with upper kHz QPO centroid
frequency $\nu_u^0$, which has only been observed previously, somewhat
marginally, in GX~17+2 \citep{homan} and on a single occasion in
4U~1728$-$34 \citep{migliari}, is in accordance with a prediction of
the relativistic precession model \citep{stella} for the behavior of
$\Delta\nu$ at low QPO frequency.  
% a = 23.3112774 
% siga =36.7459412 
% b = 0.633641183 
% sigb = 0.0985191166 
% chi2 = 2.50026989
% q = 0.980876207 
%3) Delta_nu vs nu with a constant (using plotascii)
%#Unreduced chi-squared/dof: 59.9 / 10 
%#Deltachis^2 = 1 
%#Par. value: 250.684662 + error: 7.09662294 - error: 7.09665871 
%quadratic mean error: 7.09664059 

In Fig.~\ref{rel-prec} we show a plot of the centroid frequency
separation $\Delta\nu\equiv\nu_u^0-\nu_\ell^0$ vs. $\nu_u^0$. From an
F-test for additional term we find that in our data, $\Delta\nu$ is
inconsistent with being constant at the 5.2$\sigma$ level; a linear
fit yields a line with slope 0.6$\pm$0.1.  The relation predicted by
the precession model for negligible frame dragging,
$\Delta\nu=\nu_u^0\sqrt{1-6\left(2\pi GM\nu_u^0\right)^{2/3}/c^2}$
describes our data points well ({\it drawn curve}).  The best fit to
just our Cir~X-1 points ($\chi^2/dof=0.2$) gives a mass for the
neutron star of $M=(2.2\pm0.3) M_{\sun}$.  This is high, but agrees
with inferences drawn from this model for other sources
\citep{stella}, and indeed the best-fit curve passes through the cloud
of scattered points corresponding to all other sources (which of
course can be expected to have different masses).  Note, however, that
as already discussed by \citet{stella}, for some other individual
sources no exact match to this model is possible without introducing
additional physical assumptions.  We further note, that Alfv\'en wave
oscillation models \citep{ZhangCM, rezania, ZhangCM2} which also
predict $\Delta\nu$ to increase at low $\nu_u^0$, can fit our data as
well.

As is evident from the data presented in the rightmost column of
Table~\ref{khz-table}, while the twin kHz QPO frequencies are clearly
inconsistent with a 2:3 ratio, a 1:3 ratio is closer, although still
formally inconsistent with the data at 99\% confidence.  A 1:3 ratio
corresponds to one of the resonances that have been considered by
\citet{1235}.  However, since that ratio would arise from a
combination frequency of the main 2:3 resonance, the weakness of the
main resonance (which is not significantly detected) would in such an
interpretation remain unexplained.  Beat-frequency models, based on a
constant frequency separation $\Delta\nu$ related to the spin
frequency, would have difficulties explaining the factor 2 change in
frequency separation of the kHz QPOs of Cir~X-1.  The modified
beat-frequency model proposed by \citet{lambmil} to explain the modest
{\it decrease} of $\Delta\nu$ with QPO frequency previously observed
in a number of sources would need further modification to explain the
considerable {\it increase} that we observe instead. The range of
upper kHz QPO frequencies we observe (229--505~Hz) is nearly entirely
below that expected in the sonic-point beat-frequency model
(500--1200~Hz;
\citealt{coleI}).

Although our discovery of twin kHz QPOs in Cir~X-1, as well as the
observed frequency-frequency correlations, put this system on a
footing with the neutron-star LMXBs, the fact that frequency range
over which the upper kHz QPOs are observed is mostly well below that
of other sources further confirms the exceptional nature of the
source.  These low frequencies might be taken to indicate the presence
of a massive neutron star, if for example the upper kHz QPO would
derive from the orbital frequency at some radius set by general
relativity, scaling like $1/M$.  However, when taken at face value the
fact that the observed frequencies are strongly variable makes this
interpretation unlikely.  The possibility remains that the highest kHz
QPO frequencies in a source derive through general relativity from
$M$, and the observed range of frequencies in turn derives from this
maximum.

Most proposed models interpret the strong variations of the
frequencies as due to variations in the inner radius of the Keplerian
disk flow.  This radius could be set by a number of different
mechanisms, such as electromagnetic stresses (in which case the inner
radius would be the radius of the magnetosphere), or alternatively
radiative stresses (radial radiation force and radiation drag on the
accreting matter; \citealt{coleI}).  Within the framework of such
ideas the low kHz QPO frequencies in Cir~X-1 would then imply a
relatively stronger magnetic field for the neutron star, or
alternatively a larger ratio of radiation flux density to matter
density at the inner disk edge. In the description proposed by
\citet{parlines} for the kHz QPO 'parallel tracks' phenomenon,
relatively larger radiative stresses might occur if a larger fraction
of the accretion takes place not through the (thin) disk, but in a
more radial inflow, leading to relatively more radiation, and hence
more radiative stress, being produced at the inner disk edge per gram
of accreting disk material.  The existence of such a stronger radial
inflow might possibly be related to the unusual systemic
characteristics of Cir~X-1, such as the periodic episodes of rapid
mass transfer and/or the large companion mass and likely, radius.

It is interesting to note, that a tendency towards relatively low kHz
QPO frequencies is already known to be one of the characteristics
distinguishing Z sources from other neutron star LMXBs.  As noted
above, Cir~X-1 shares several more characteristics with the Z sources.
So, Cir~X-1 may possess the same physical property that makes Z
sources stand out, but more extreme.  If that property would
be a stronger magnetic field, as has been discussed before
\citep[e.g., ][]{psalt95}, then additional ingredients (such as smearing of
rapid variability by scattering in material surrounding the neutron
star) are required to explain why no coherent pulsations are observed.
If instead a stronger \hbox{(radial flow / disk flow)} ratio causes the
Z-source phenomenon (possibly, like in Cir~X-1, related to systemic
characteristics such as a larger companion radius), then the radiative
transfer of the emerging radiation through this inflow may be what
causes the spectral effects (the Z track), and the larger inner disk
radius what causes some of the timing effects (lower kHz QPO
frequencies) that set these sources apart.

\appendix

\section{The determination of the Poisson noise spectrum}\label{appendix}

For a high-frequency analysis an accurate description of the
deadtime-modified Poisson noise power spectrum is required. In the
RXTE/PCA this spectrum is mainly determined by the values of the event
deadtime $t_d$ and the VLE-window\footnote{In our data only the 3rd
VLE window setting appears (a.k.a. setting 2; $\sim$150~$\mu$s) and
for that reason this Appendix only refers to that setting.}
$t_{VLE}$.  The technical descriptions of RXTE give only approximate
values for these parameters, $t_d$=10 and $t_{VLE}$=150~$\mu$s,
respectively \citep{zhang2}; values of 8.5$\mu$s have also been used
for $t_d$ (\citealt{td1}, \citealt{marc}, \citealt{td2}), while a
method using additional small shifts was introduced by \citet{marc} to
compensate for deviation from the observations.  Recently, a
calibrated VLE window value of $170\mu$s was reported by
\citet{jahoda}.  Since these values could in principle vary and in
some cases depend on source spectrum, and since the values mentioned
above do not agree with our observations to high precision (see
Fig.~\ref{PL}), we instead estimated the deadtime parameters directly,
making good use of the large sample of data available for Cir~X-1.

We used 84 representative RXTE observations of Cir~X-1 distributed
over all epochs and with count rates between 300 and 6000~c/s/PCU.  No
dependencies of the deadtime parameters on epoch or count rate were
found. Observations with high time resolution were preferred but
lower-resolution ones, as well as observations with extremely high
resolution (Nyquist frequency $\sim$33~kHz) all gave consistent
results.  In order to avoid power contributions from known source
variability components, we looked only at frequencies above 1.6 kHz
(up to the Nyquist frequency).  All observations used for this
analysis covered the whole energy range (channels 0--249).  

We used the function approximately describing the Poisson power
spectrum according to \citet{zhang,zhang2}:
\begin{eqnarray*}
P_{\nu} = 2 - 4 r_0 t_d \left(1 - \frac{t_d}{2t_b}\right) -
2\frac{N-1}{N} r_0 t_d \left(\frac{t_d}{t_b}\right) \cos{2\pi\nu t_b}+
2r_{VLE} r_0 t^2_{VLE} \left(\frac{\sin{\pi t_{VLE}\nu}}{\pi t_{VLE}\nu}\right)^2, 
\end{eqnarray*} 
where $t_d$ and $t_{VLE}$ were defined above, $N$ is the number of
points in the time series, $r_0$ is the count rate per detector,
$r_{VLE}$ is the VLE count rate per detector, and $t_b$ is the time
bin width, so that $1/2t_b$ is the Nyquist frequency.  By
using $t_d$ and $t_{VLE}$ as free parameters and fixing the other ones
to their known values, we fitted this function to the power spectra
above 1.6 kHz for each of the 84 observations. All fits represented
the data well, without need for shifting, for a $t_d$ value spanning
only a narrow range, while $t_{VLE}$ was also well determined. Based
on these measurements of the deadtime parameters and their
distributions we adopted the values $t_d$=8.87$\mu$s and
$t_{VLE}$=162$\mu$s to use throughout our entire analysis.
%The average values and their one sigma errors are \begin{eqnarray*}
%t_d=8.87\pm0.2\mu\mbox{s},\;\;\; t_{VLE}=162^{+7}_{-12}\mu\mbox{s}
%\end{eqnarray*} \rem{what was chi2 for these averages?  do we have the
%info??? 8.97!}
These numbers are in the range of values given by the RXTE team
\citep{jahoda} and used by various other authors
\citep{td1,td2}. Whether the above values accurately describe the
deadtime values for other observations needs to be tested using
sources with other spectral shapes and flux levels, as well as for the
other VLE settings. 

As a test we compared our fit results to those we would have obtained
with other choices for the deadtime modified Poisson spectrum
calculation.  Deadtime parameters $t_d$=10$\mu$s, $t_{VLE}$=170$\mu$s
gave a very bad match to many of our power spectra and in those cases
produced no reasonable fits, but some previously used deadtime
parameter values (near $8.5,150\mu s$) gave reasonable results.  We
refitted the power spectra for the observations showing twin kHz QPOs
(Section~\ref{twin}) using $8.5$ and $150\mu$s and found no systematic
deviations in the kHz QPO frequencies and coherences compared to
fitting with our adopted values.  This was also the case when using
the method of shifting the Poisson noise spectrum to better match the
power spectra at high frequencies.  We did find that our adopted
method tended to be slightly more conservative with respect to
detecting high-frequency features than the shifting method, where
occasionally an extra component was necessary at high frequencies
which was not significant with our method (e.g., power spectrum E of
Fig.~\ref{fits}).  In any case, twin kHz QPOs reported as significant
herein were also significant with these two previously used methods of
estimating the deadtime modified Poisson noise spectrum.

\clearpage

\begin{table}
\caption{
Root observation IDs of the RXTE observations used in
    this analysis.}%not in time-order} \
    \label{obsids}
    \begin{center}
    \begin{tabular}[t]{l} \hline 10068-08, 10122-01, 10122-02,
    10122-03\\ 20094-01, 20095-01, 20097-01, 20415-01\\ 30080-01,
    30081-01\\ 40059-01\\ 50136-01\\ 60024-01, 60025-01\\ 70020-01,
    70020-02, 70020-03, 70021-01\\ 80027-01, 80114-01\\ 90025-01,
    90426-01\\ \hline \end{tabular}
    \end{center}
\end{table} 

%\clearpage

\begin{table}
\scriptsize
    \caption{
%\scriptsize 
All observations showing twin kHz QPOs with significances
    of more than 3$\sigma$ for both kHz QPOs (A--G), and more than
    3$\sigma$ for the lower kHz QPO and at least 2.5$\sigma$ for the
    upper kHz QPO (H--K).  Listed are the observation ID, centroid
    frequencies, rms amplitudes, and quality factors, as well as the centroid 
    frequency difference and ratio. }  
    \label{khz-table}
\begin{center}
%    \begin{ruledtabular}
    \begin{tabular}[t]{cccccccccc}
Data &      Obs-ID   & $\nu_{\ell}^0$ & rms           & Q           &$\nu_u^0$     &   rms & Q & $\Delta\nu$ & $\nu_u^0/\nu_\ell^0$ \\
Set  &               & (Hz)           & (\%)          &             &(Hz)          & (\%)    &        \\
\hline
A&10122-03-08-00 &  56.1$\pm$1.3  & 3.88$\pm$0.11 & 0.8$\pm$0.1 &229$\pm$18 & 1.41$\pm$0.24 & 2.1$\pm$1.2 &173$\pm$18&4.1$\pm$0.3\\
B&10122-01-15-01 &  83.1$\pm$2.5  & 2.94$\pm$0.15 & 1.3$\pm$0.2 &281$\pm$32 & 1.69$\pm$0.32 & 1.6$\pm$0.9 &198$\pm$32&3.4$\pm$0.4\\
C&20095-01-08-00 &  97.5$\pm$2.7  & 3.13$\pm$0.12 & 1.0$\pm$0.1 &311$\pm$20 & 1.93$\pm$0.21 & 1.8$\pm$0.5 &214$\pm$20&3.2$\pm$0.2\\
D&20097-01-32-00 &  100$\pm$26    & 2.74$\pm$0.40 & 0.8$\pm$0.5 &419$\pm$48 & 2.03$\pm$0.43 & 2.6$\pm$1.6 &319$\pm$54&4.2$\pm$1.2\\
E&10122-01-18-00 & 106.5$\pm$2.5  & 2.93$\pm$0.13 & 1.2$\pm$0.1 &359$\pm$42 & 2.28$\pm$0.25 & 0.9$\pm$0.4 &252$\pm$42&3.4$\pm$0.4\\
F&10068-08-01-00 & 122.3$\pm$7.2  & 2.61$\pm$0.19 & 0.9$\pm$0.2 &407$\pm$33 & 1.74$\pm$0.24 & 1.8$\pm$0.7 &284$\pm$34&3.3$\pm$0.3\\
G&20094-01-01-030& 127.6$\pm$3.6  & 2.53$\pm$0.12 & 1.0$\pm$0.1 &445$\pm$20 & 1.60$\pm$0.24 & 1.8$\pm$0.5 &317$\pm$20&3.5$\pm$0.2\\
H&20094-01-01-01 & 136.5$\pm$5.4  & 2.65$\pm$0.21 & 1.1$\pm$0.2 &404$\pm$17 & 2.08$\pm$0.19 & 1.9$\pm$0.5 &267$\pm$18&3.0$\pm$0.2\\
I&20095-01-11-00 & 137.4$\pm$5.4  & 2.19$\pm$0.39 & 1.1$\pm$0.3 &438$\pm$38 & 1.47$\pm$0.31 & 2.0$\pm$1.3 &301$\pm$29&3.2$\pm$0.3\\
J&20094-01-01-020& 157$\pm$22     & 2.37$\pm$0.26 & 0.6$\pm$0.2 &497$\pm$61 & 1.47$\pm$0.33 & 1.3$\pm$0.6 &340$\pm$64&3.2$\pm$0.6\\
K&20094-01-01-02 & 226$\pm$18     & 1.24$\pm$0.28 & 1.8$\pm$1.0 &505$\pm$51 & 1.65$\pm$0.22 & 1.5$\pm$0.5 &279$\pm$54&2.2$\pm$0.3\\
%30081-06-03-00$^{\ast}$&  14.8$\pm$0.2  & 3.17$\pm$0.35  & 1.2$\pm$0.1 & 18.4$\pm$ 3.3 & 3.95$\pm$0.33 & 0.4$\pm$0.1 &&\\
%30081-06-03-03$^{\ast}$&  14.6$\pm$ 0.4  & 2.46$\pm$0.58 & 1.6$\pm$0.5 & 13.1$\pm$ 3.1 & 4.57$\pm$0.39 & 0.3$\pm$0.1 &&\\
%($\hookrightarrow \nu_3$)& 277.8$\pm$7.1  &  0.87$\pm$0.23   & 5.5$\pm$5.3 & & &\\
\hline
    \end{tabular}
%    \end{ruledtabular}
  \end{center}
\end{table}

\clearpage

\begin{figure} \includegraphics[scale=0.9]{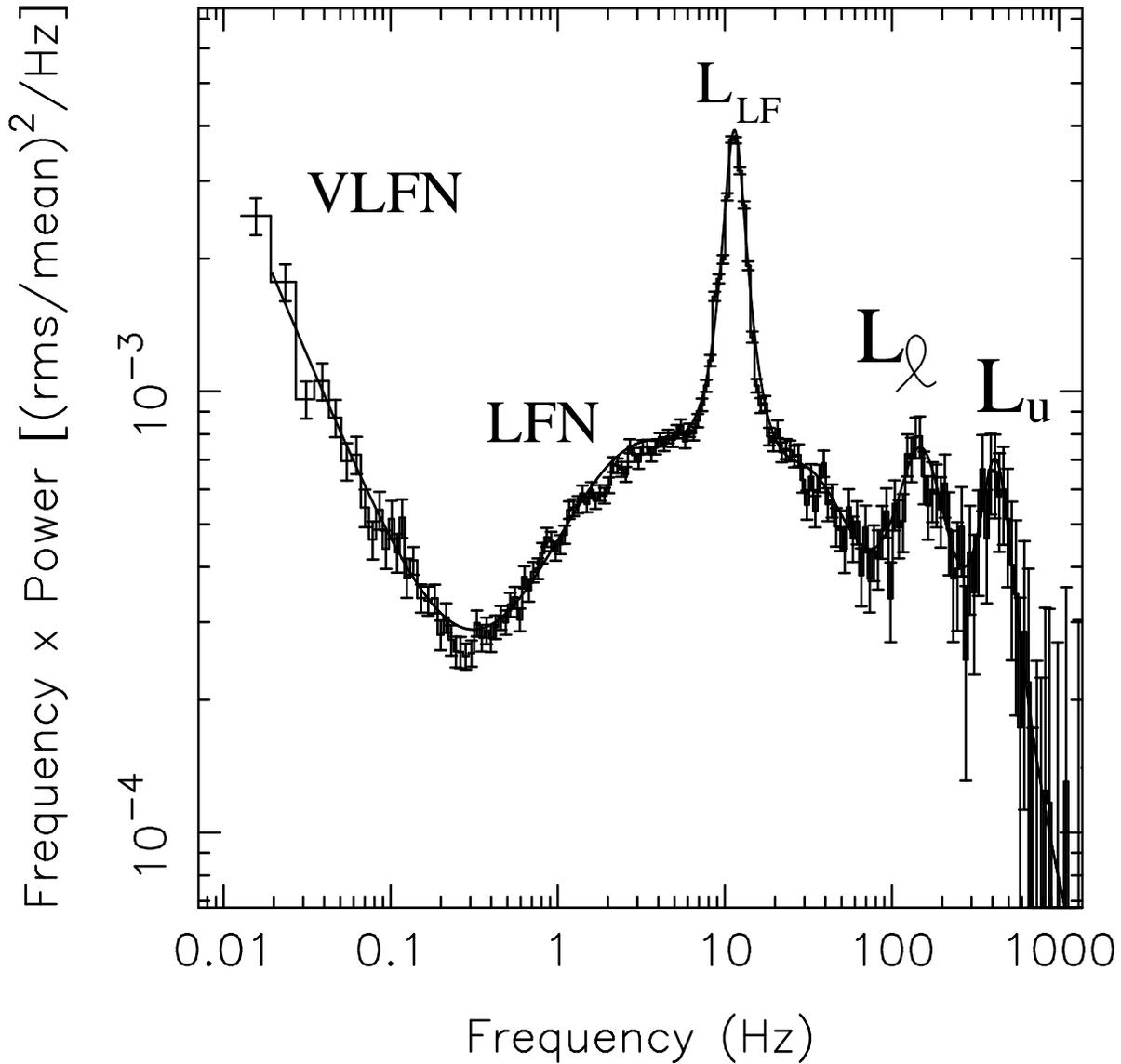}
\caption{\label{ps} The power spectrum and fit function of observation
20094-01-01-01 in a power times frequency representation.  The lower
and upper kHz QPOs ($L_\ell$ and $L_u$) and the low-frequency QPO
($L_{LF}$) are marked, as well as are the Low Frequency Noise (break
component) and Very Low Frequency Noise (power law at low
frequency). The (unmarked) component present above the low frequency
QPO (here at $\sim$22~Hz) is usually consistent with being a harmonic
to $L_{LF}$, particularly when it has high $Q$,
cf. Fig.~\ref{twoLF}. } \end{figure}

\clearpage

\begin{figure}
\centering 
\includegraphics[scale=1.0]{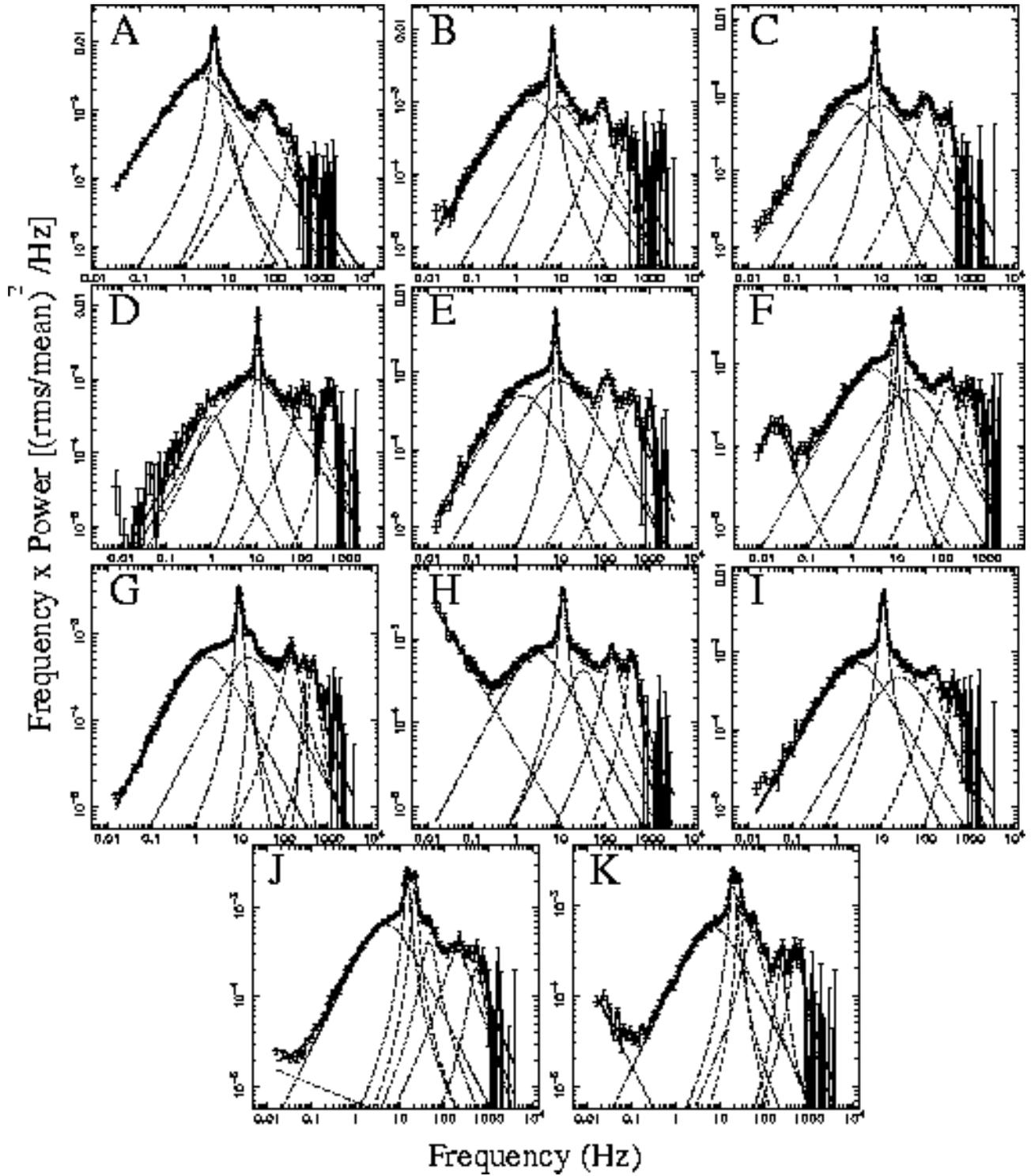}
\caption{Power spectra and fit functions in the power spectral density times
frequency representation for the 11 observations showing twin kHz
QPOs.  The curves mark the individual Lorentzian components of the
fit.}
\label{fits}
\end{figure}

\clearpage

\begin{figure}
\begin{center}
\includegraphics[angle=-90,scale=0.675]{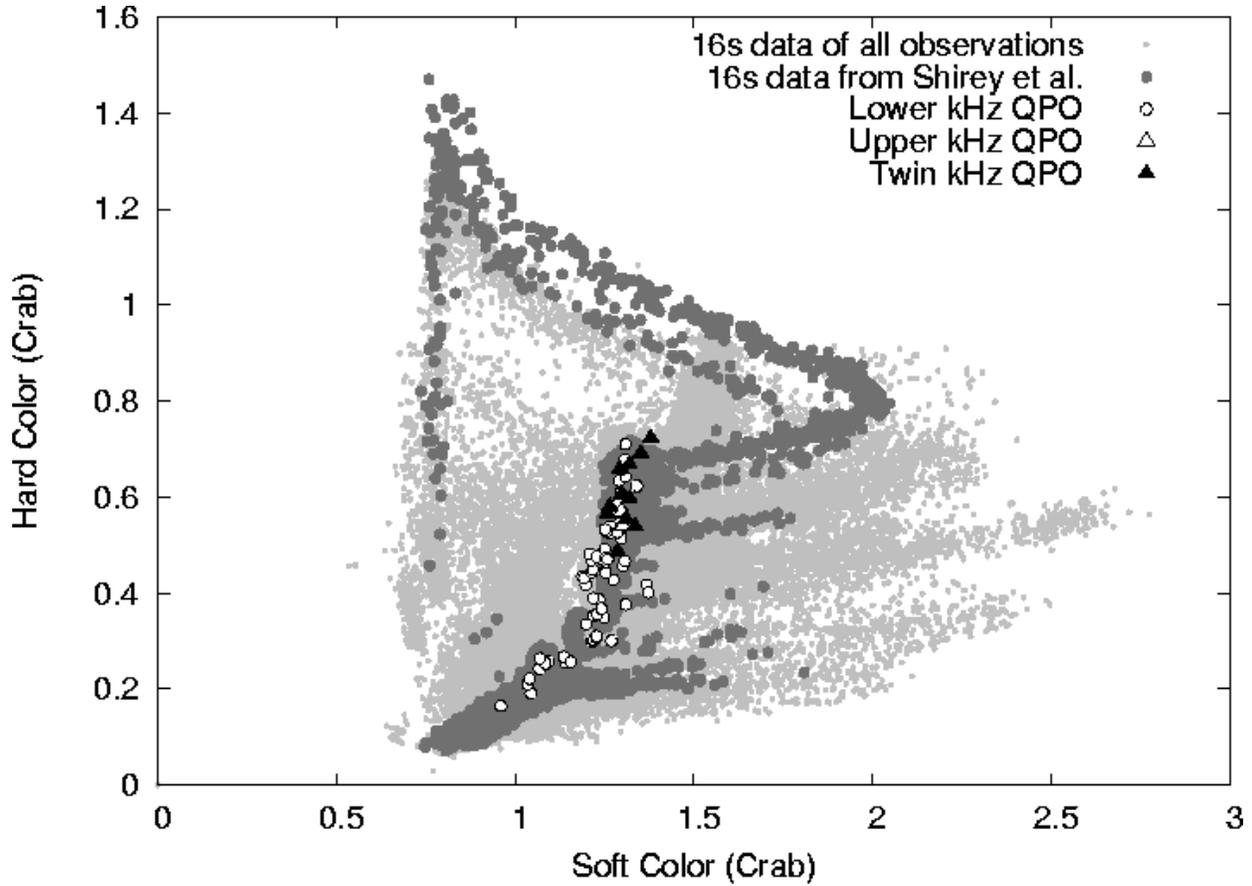}
\end{center}
\caption{\label{ccd}
The color-color diagram of all observations of Cir~X-1 that we used in
16-s time resolution ({\it grey}) with the data already reported by
\citet{shirey} plotted in a darker shade.  Superimposed are the
average colors for each of the 80 observations with kHz QPOs ({\it
white and black}); see plot for definition of the symbols.  Hard color
bands are 8.5--18 keV/4.8--8.5 keV, soft color bands 3.4--4.8
keV/2--3.4 keV.}
\end{figure}

\clearpage

\begin{figure}
\begin{center}
\includegraphics[angle=-90,scale=0.675]{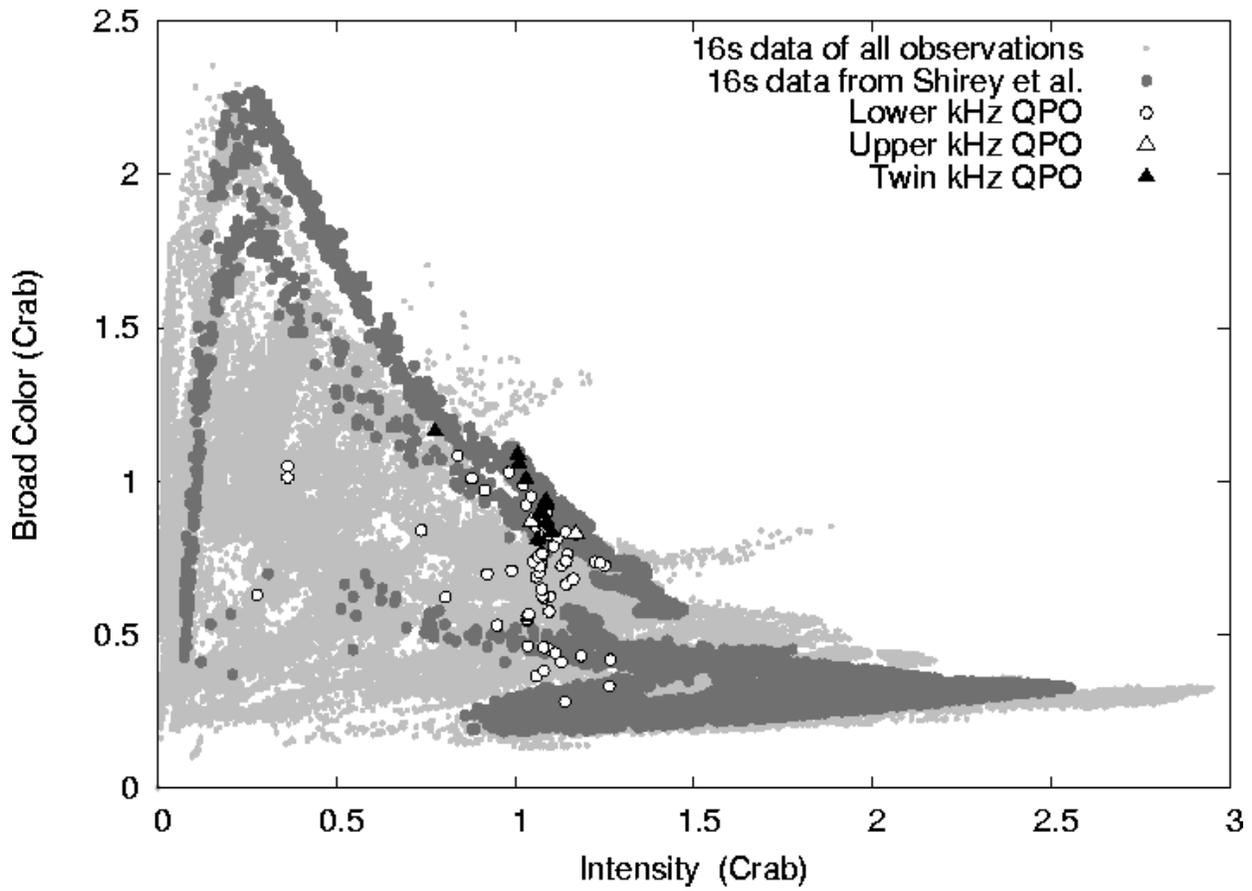}
\end{center}
\caption{\label{hid}
Hardness-intensity diagram. All symbols have the same meaning as in
Fig.~\ref{ccd}.  Broad color bands are 6.3--13/2--6.3~keV, intensity
is 2--18~keV count rate.}
\end{figure}

\clearpage

\begin{figure}
\begin{center}
\includegraphics[angle=-90,scale=0.675]{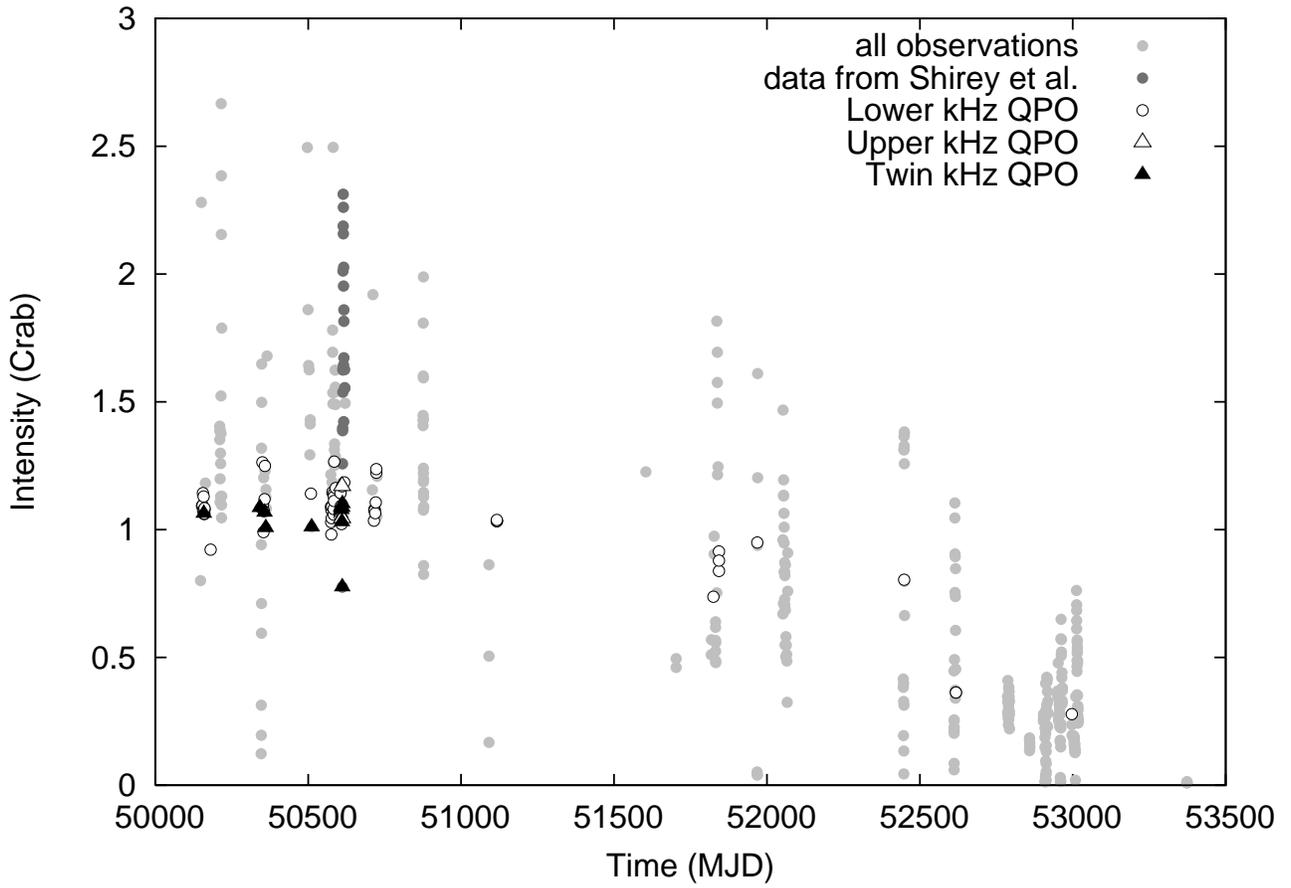}
\end{center}
\caption{\label{lc}
Long term light curve of Cir~X-1.  Each data point represents the
average of one of our 497 observations. Apart from that, all symbols
have the same meaning as in Fig.~\ref{ccd}.}
\end{figure}

\clearpage

\begin{figure}
\begin{center}
\includegraphics[angle=-90,scale=0.675]{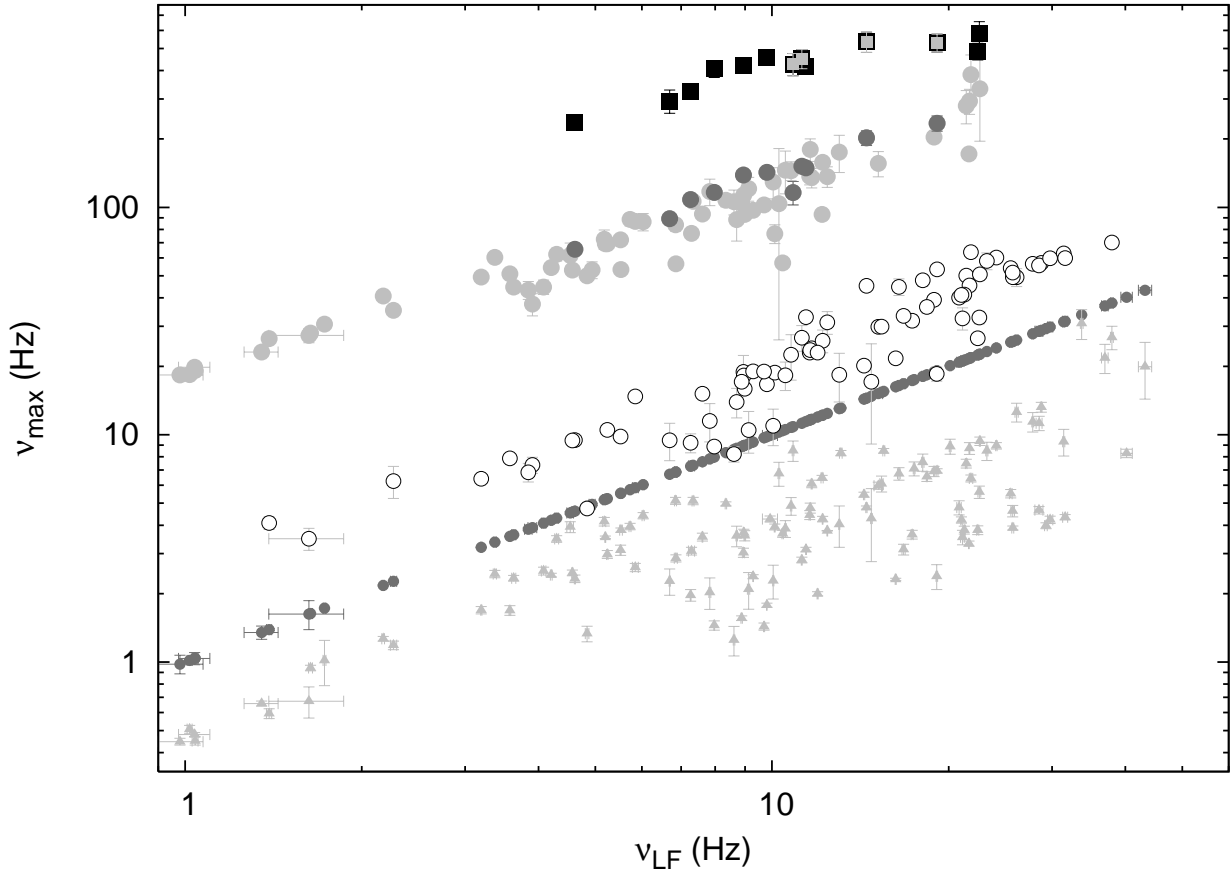}
\end{center}
\caption{\label{fr-fr}
All measured characteristic frequencies as a function of low-frequency
QPO frequency $\nu_{LF}$.  Plotted from top to bottom are the
components identified as: $L_u$ ({\it squares}), $L_\ell$ ({\it large
filled circles}), the component ({\it open circles}) that is usually
dominated by the harmonic to $L_{LF}$, $L_{LF}$ ({\it small filled
circles}), and the low frequency noise or break component ({\it
triangles}).  The 11 lower kHz QPO points that appear as one of a twin
kHz QPO are plotted with a darker shade, and the 4 upper kHz QPOs that
are only 2.5--3$\sigma$ significant with a lighter-centered
square. Error bars are plotted but are often smaller than the
symbols.}
\end{figure}

\clearpage

\begin{figure}
\begin{center}
\includegraphics[scale=0.675,angle=-90]{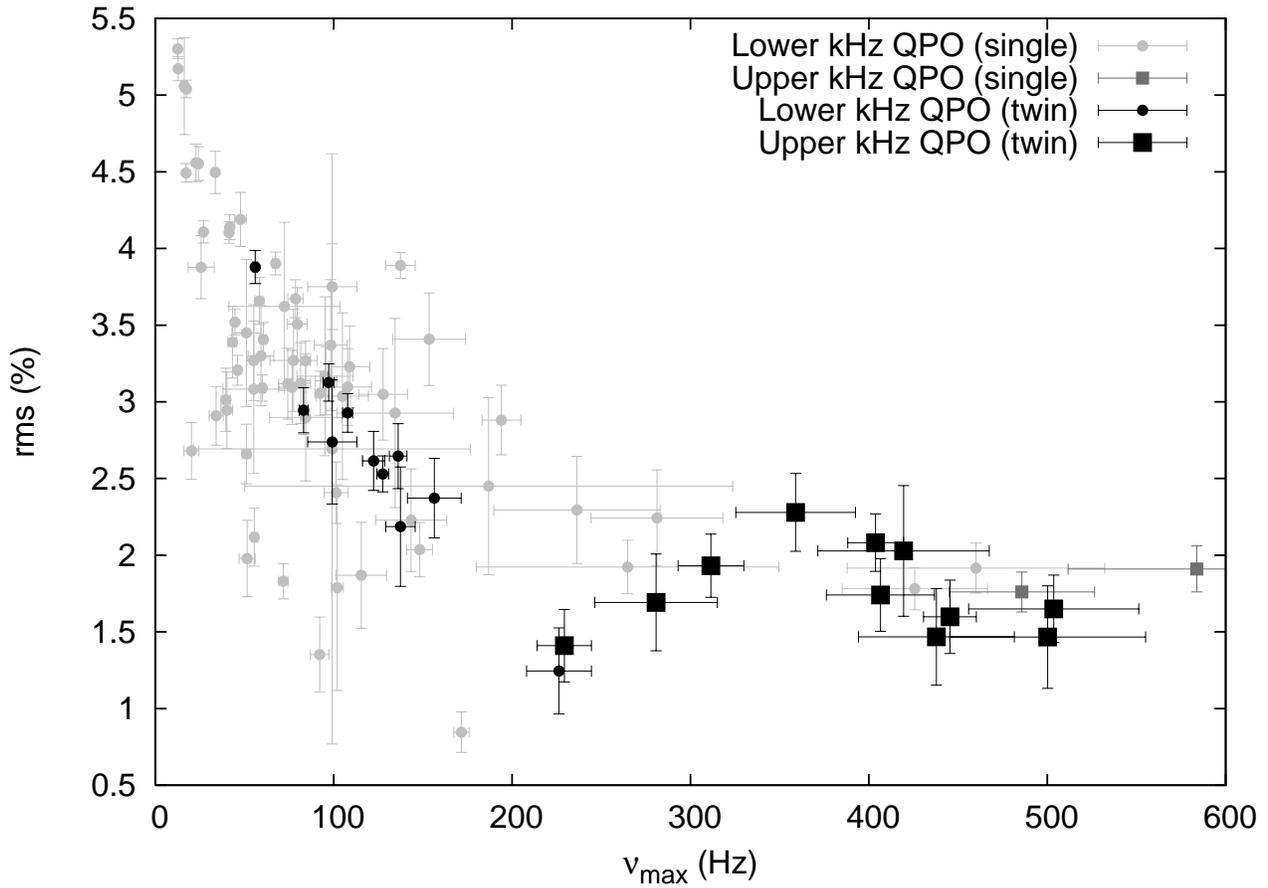}
\end{center}
\caption{\label{rms} The fractional rms amplitude of each kHz QPO
as a function of its own characteristic frequency.  Single and twin
lower and upper kHz QPOs are plotted with different symbols as
indicated in the figure.}
\end{figure}

\clearpage

\begin{figure}
\begin{center}
\includegraphics[scale=0.675,angle=-90]{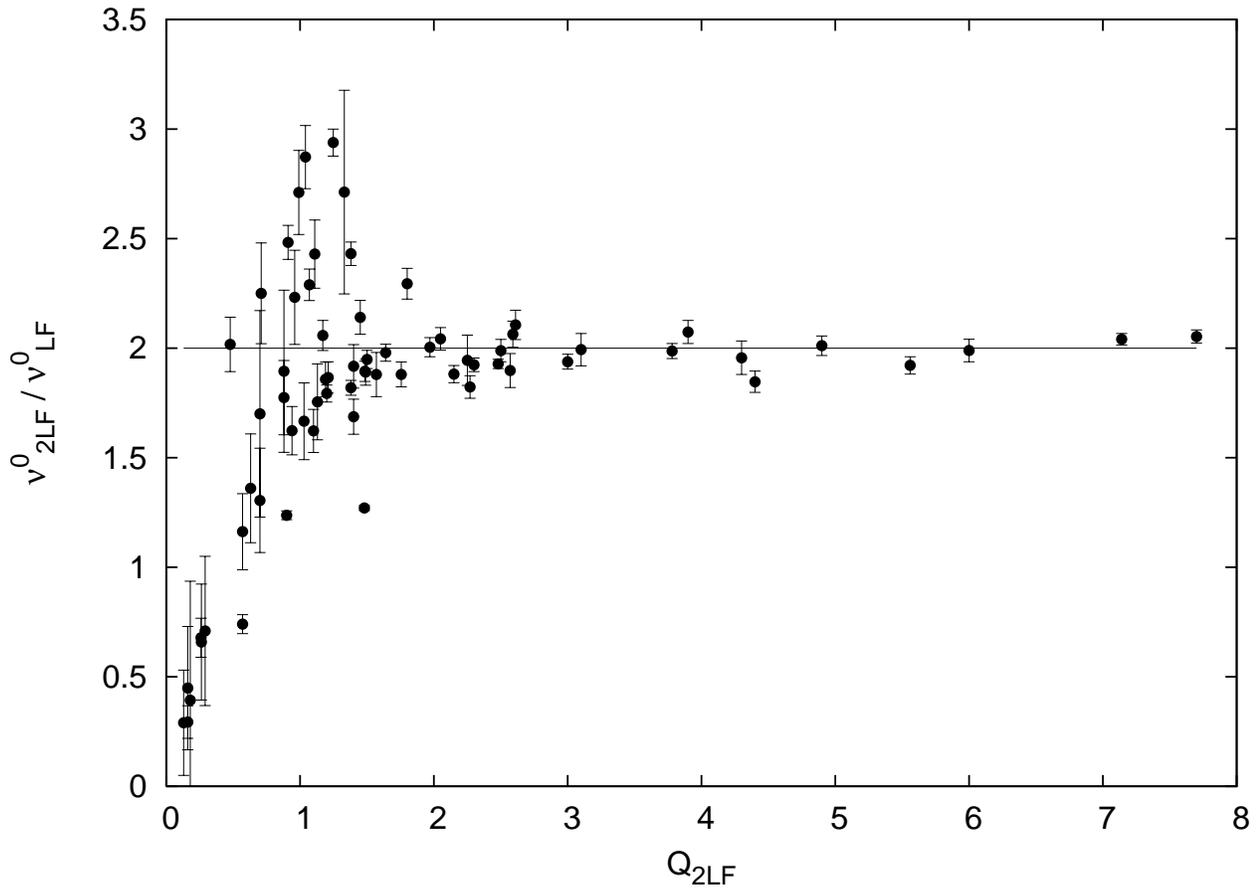}
\end{center}
\caption{\label{twoLF}
The ratio of the centroid frequency of the component above $L_{LF}$ to
the centroid frequency of $L_{LF}$ vs. the coherence of the former.
Clearly, when the coherence is good, this component is at twice the
frequency of $L_{LF}$.}
\end{figure}

\clearpage

\begin{figure}
\begin{center}
\includegraphics[scale=0.675,angle=-90]{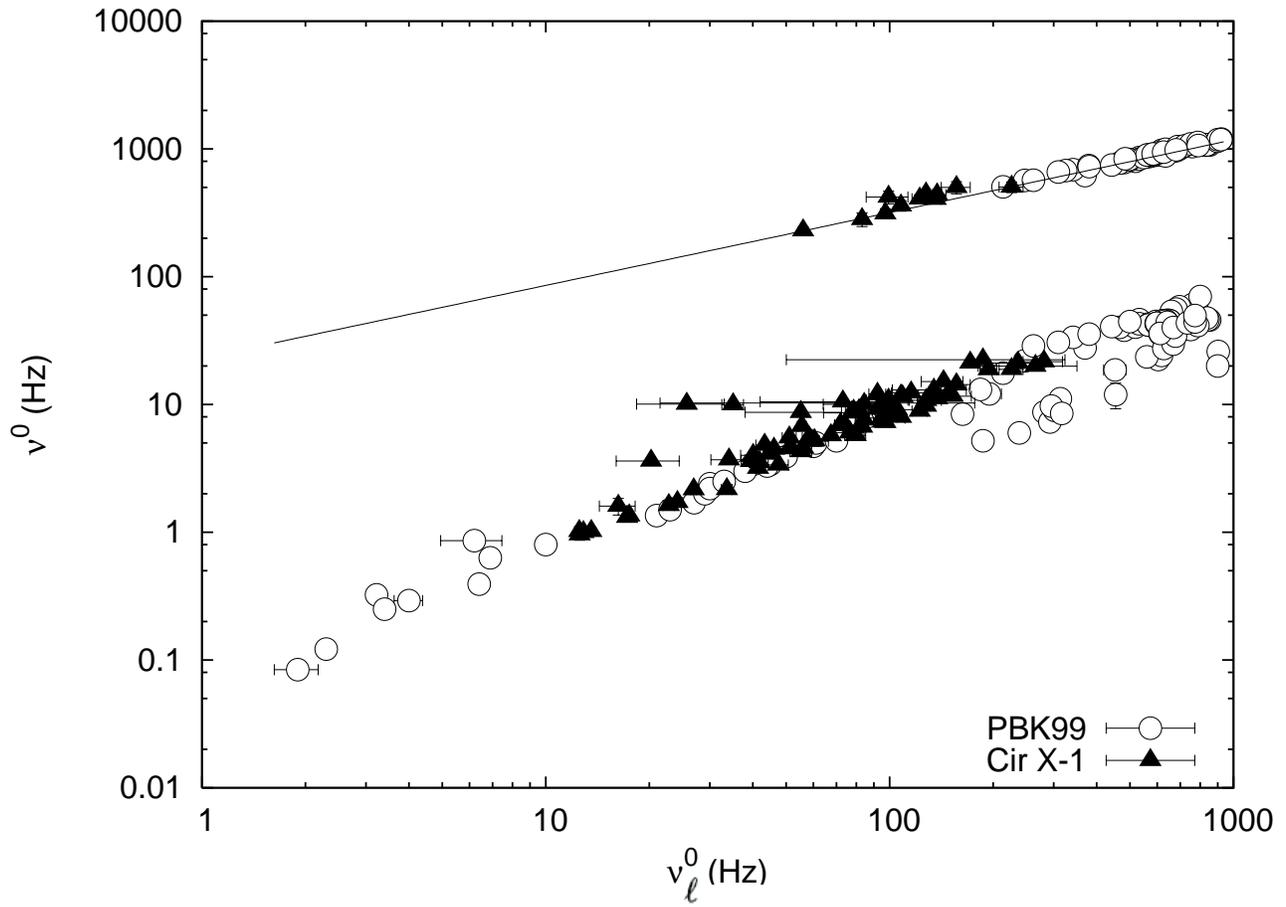}
\end{center}
\caption{\label{pbk}The centroid frequency of the upper kHz QPO and 
the LF QPO vs. that of the lower kHz QPO for Cir~X-1 ({\it black triangles})
and other sources used by \citet{pbvdk} ({\it open circles}).  The line
is a power law fit to all upper kHz QPO points.}
\end{figure}

\clearpage

\begin{figure}
\begin{center}
\includegraphics[scale=0.75]{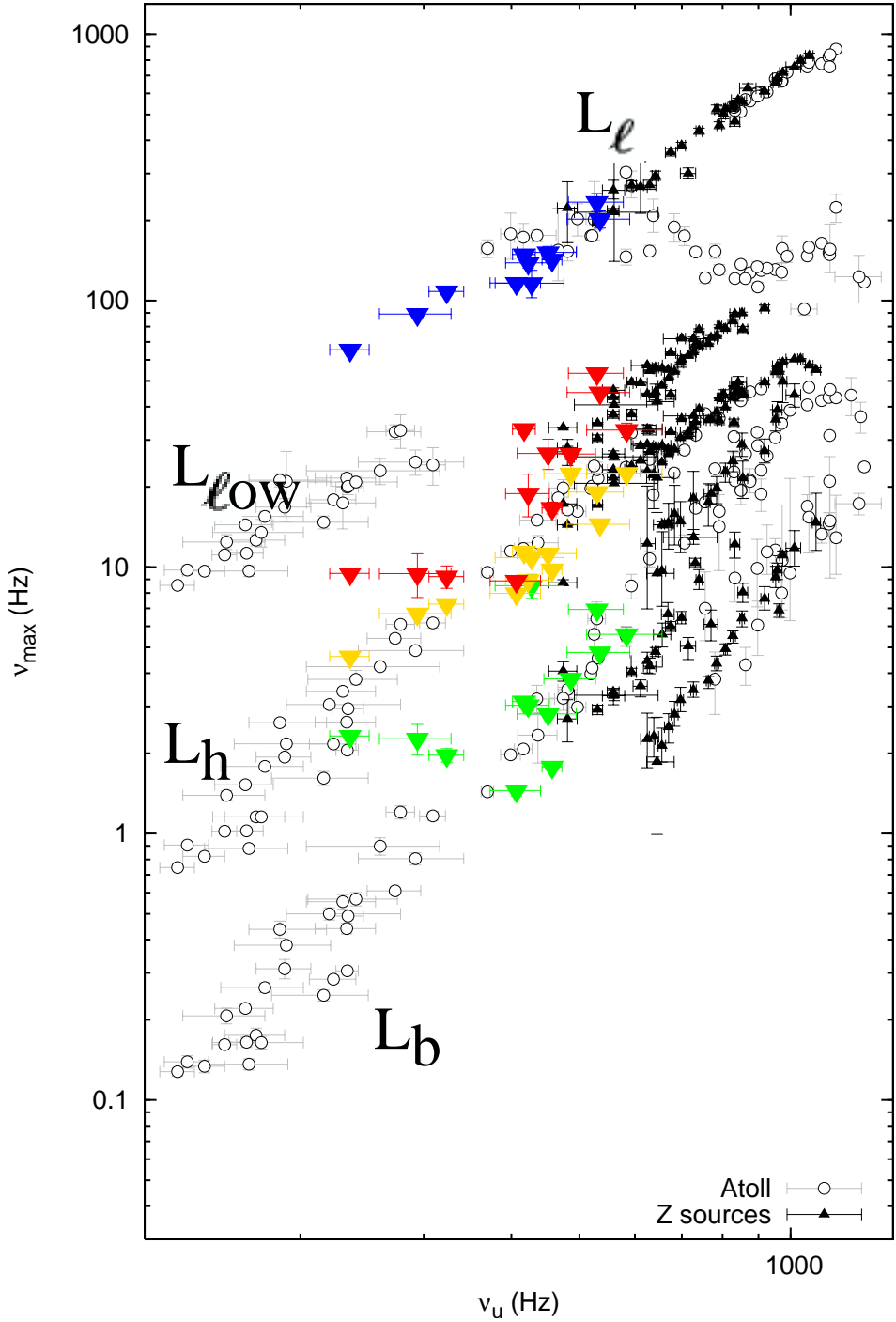}
\end{center}
\caption{\label{vstrfig} Characteristic frequencies of all components
vs. the characteristic frequency of $L_u$.  Previous observations of Z
and atoll sources are plotted in black as indicated in the figure; the
tracks of $L_\ell$, $L_{\ell ow}$, $L_h$ and $L_b$ as previously
identified in atoll sources are indicated.  Colors distinguish the
different components we identified in Cir~X-1 (cf. Figs.~\ref{ps} and
\ref{fr-fr}): lower kHz QPO ({\it blue}), LF QPO harmonic ({\it red}),
LF QPO ({\it orange}), low-frequency noise/break component ({\it
green}).}
\end{figure}

\clearpage

\begin{figure}
\begin{center}
\includegraphics[scale=0.675,angle=-90]{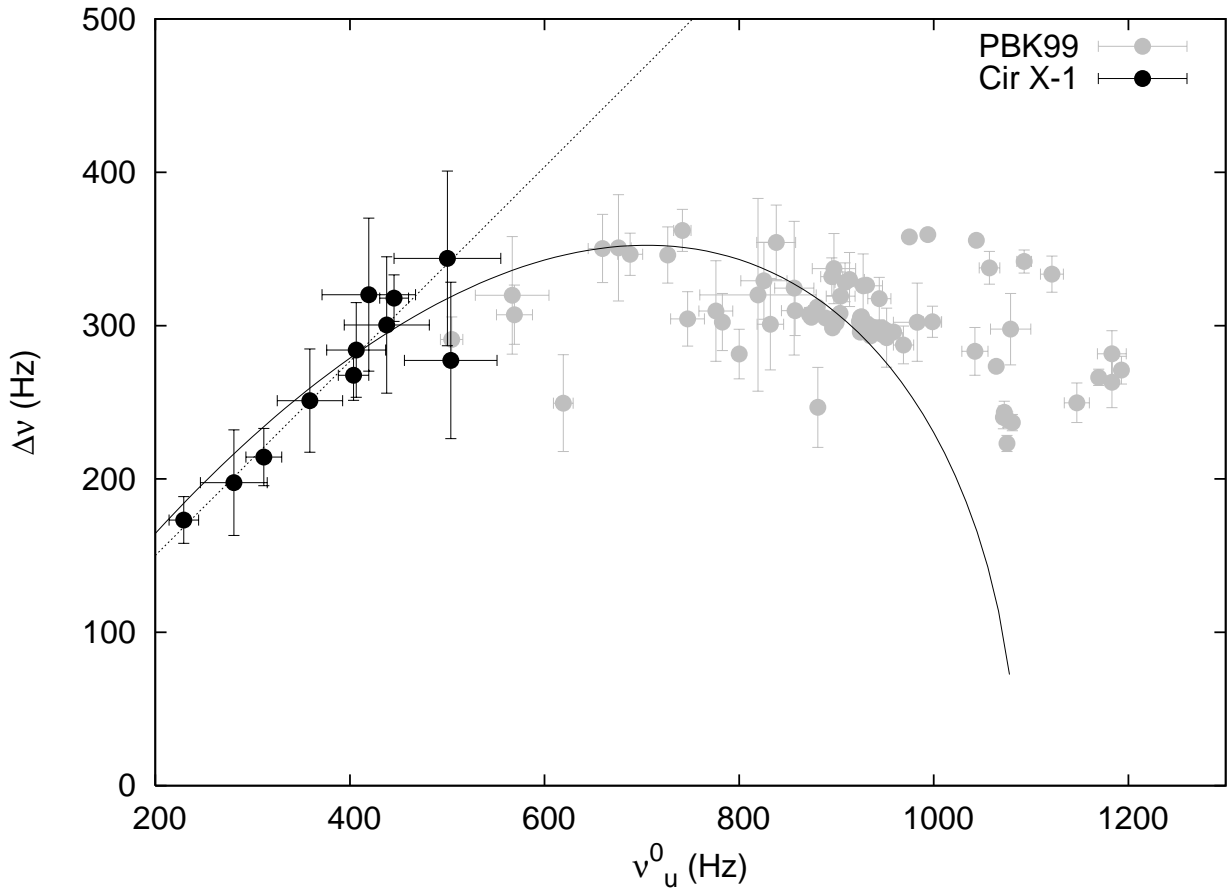}
\end{center}
\caption{\label{rel-prec} The centroid frequency separation $\Delta\nu$ of the
twin kHz QPOs vs. the upper kHz QPO centroid frequency.  Also plotted
%are the straight line ({\it dashed}) as well as the curves for the relativistic
%precession model ({\it drawn}) and the Alfv\'en wave model ({\it
%dotted}) best fitting the Cir~X-1 data. }
are the straight line ({\it dashed}) and the curve for the relativistic
precession model ({\it drawn}) best fitting the Cir~X-1 data. }
\end{figure}

\clearpage

\begin{figure}
\begin{center}
\includegraphics[scale=0.45]{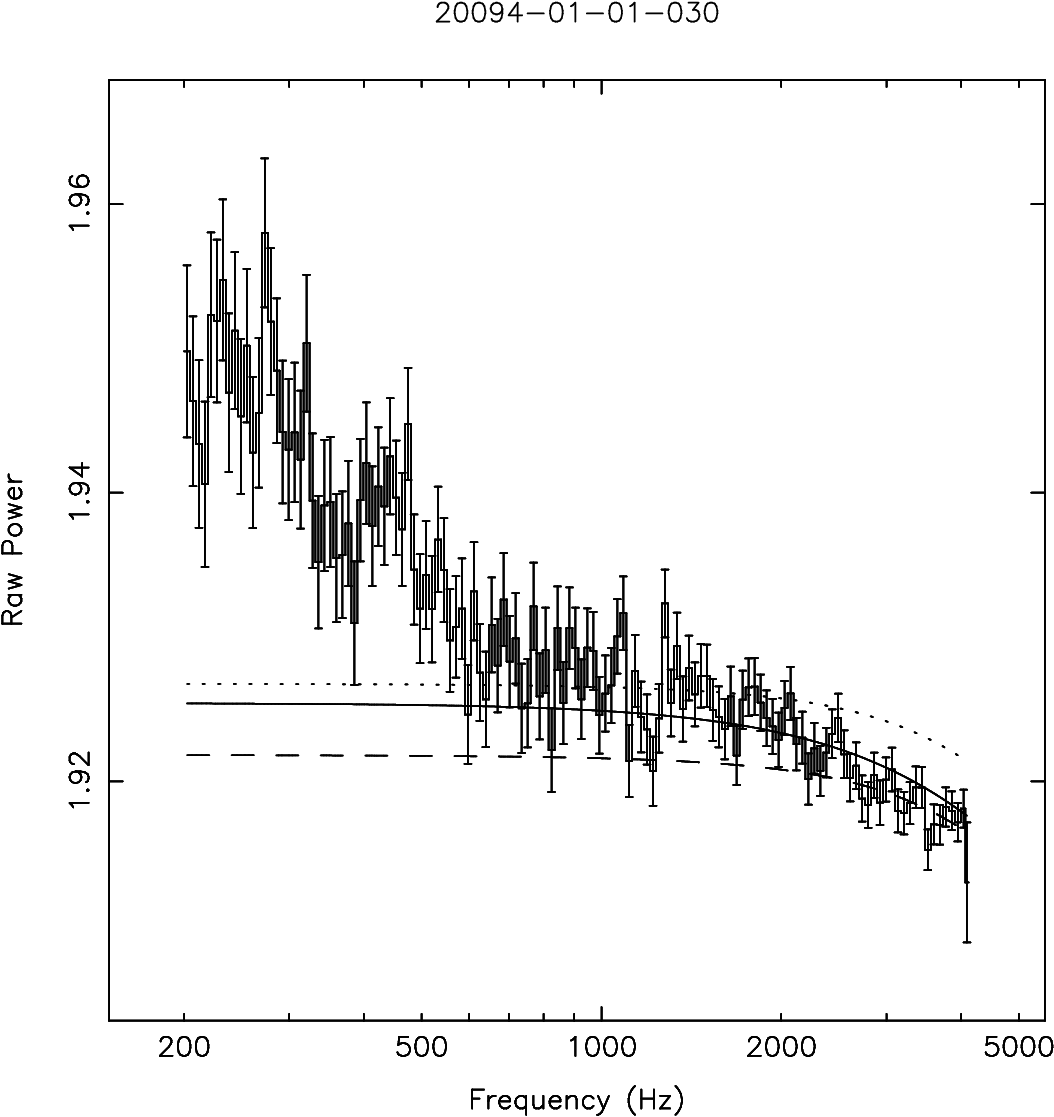}
\includegraphics[scale=0.45]{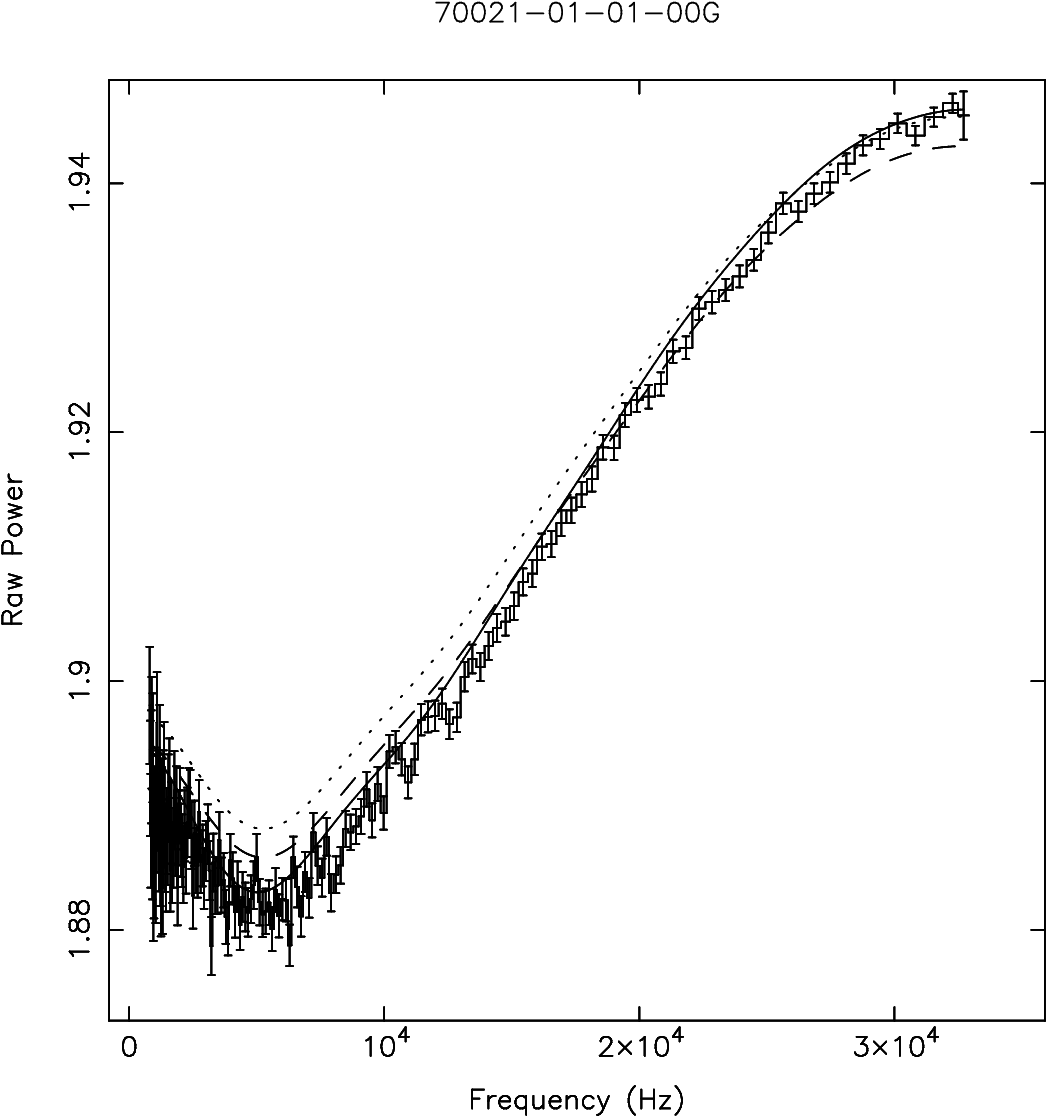}
\end{center}
\caption{\label{PL} Two examples of observations whose Poisson noise spectrum
would have been estimated less accurately with other methods. \newline
{\it Left:} The Poisson spectrum according to the model
of \citet{zhang,zhang2} using some previously suggested deadtime values
($t_d$=8.5$~\mu$sec, $t_{VLE}$=150~$\mu$sec) ({\it dotted}) fails to
precisely describe the power at high frequencies.  With a shift \citep{marc} to match
the $>0.75\nu_{Nyquist}$ region ({\it dashed}),
power at frequencies up to 2~kHz might be overestimated.
%even the third kHz QPO (see subsection \ref{twin}) would exceed the $3\sigma$ significance level.
The Poisson spectrum computed with the deadtime values adopted here
and no shifting ({\it drawn}) follows the data better in the range
$>1600$~Hz where no contribution from the source is expected, and is
more conservative with respect to kHz features than the shifted power
spectrum. \newline {\it Right:} The strong frequency dependence of the
Poisson-noise power at very high frequencies (observed for high time
resolution) cannot be described with the Zhang function ({\it dotted})
for the deadtime values mentioned above.  The curve lies
systematically above the data for several kHz, or partially above and
partially below it when shifted ({\it dashed}), at frequencies where
the data should purely be described with Poisson noise.  Again the
estimate with our adopted deadtime parameters ({\it drawn}) is better
and appears slightly conservative with respect to possible
high-frequency features.}
\end{figure}

\end{document}